\documentclass[aps,twocolumn,showpacs,floatfix]{revtex4}
\usepackage{epsfig}
\usepackage{amsfonts}

\begin{document}

\title {Electron-electron interaction correction and magnetoresistance in tilted fields in Si-based 2D systems}

\author{A.\,Yu.~Kuntsevich$^a$\thanks {e-mail:
alexkun@lebedev.ru}, L.\,A.~Morgun$^a$, V.\,M.~Pudalov$^{a,b}$}

\affiliation{$^a$ P.\,N.~Lebedev Physical Institute of the Russian Academy of Sciences, 119991 Moscow, Russia\\
$^b$ Moscow Institute of Physics and Technology,  Dolgoprudny, 141700, Russia\\}

\begin{abstract}We study diffusive electron-electron interaction correction to conductivity
by analyzing simultaneously  $\rho_{xx}$ and $\rho_{xy}$ for disordered 2D electron systems  in Si
in tilted magnetic field. Tilting the
field is shown to be a straightforward tool to disentangle spin and orbital effects. In particular, by changing the tilt angle we prove experimentally that in the field range $g\mu_BB>k_BT$ the correction depends on modulus of magnetic field rather than on its direction, which is expected for a system with isotropic $g$-factor.
In the high-field limit the correction behaves as $\ln (B)$, as expected theoretically
(Lee, Ramakrishnan, Phys. Rev. B{\bf 26} , 4009 (1982)).
Our data prove that the diffusive electron-electron interaction correction to conductivity is not solely responsible for the huge and temperature dependent magnetoresistance in parallel field, typically observed in Si-MOSFETs.
\end{abstract}
\pacs{72.20.Fr, 73.20. Jc, 73.40.Qv, 73.43.Fj}
\maketitle

\section{INTRODUCTION}
A diffusive electron-electron interaction correction (EEC) to the conductivity was predicted theoretically~\cite{altshuleraronovlee} about 30 years ago. In 2D system it is proportional to $\ln (\hbar/k_BT\tau)$
 ($\tau$ is the momentum relaxation
time) and grows in amplitude as temperature decreases. A way to experimentally single-out EEC among other numerous effects is based on its property not to affect Hall component of magnetoconductivity tensor $\sigma_{xy}$ in perpendicular magnetic field~\cite{altshulerhall}. EEC therefore gives birth to temperature-dependent and parabolic with field
contribution to the diagonal magnetoresistance $\rho_{xx}(B)$ and correction to the Hall coefficient $\rho_{xy}/B$, both being proportional to $\ln (\hbar/k_BT\tau)$.
The predicted features were observed in numerous experiments, mainly with n-type GaAs-based 2D systems~\cite{paalanen,taboryski,choi,poirier}. However, the quantitative level of agreement between theory and experiment was achived only in the 2000s by Minkov group~\cite{minkoveefirst} from simultaneous analysis of both Hall and diagonal components of resistivity tensor. The suggested method was later approbated by others~\cite{liballistictodiffusion,renardballistictodiffusion,Goh}. We note, that  Zeeman splitting effects were negligible in most of the studied systems.

Zeeman splitting was predicted to decrease the EEC value~\cite{LeeRamakr}, the physical interpretation
of this effect introduced later in Ref.~\cite{ZNABPar} consists in decreasing the effective number of triplet channels with field.
For the diffusive regime $k_BT\tau/\hbar\ll 1$, the EEC is predicted to be quadratic-in-field in the low field limit,  and proportional to logarithm of field in the high field limit.

 Experimentally, however, the effect of Zeeman splitting on EEC in the diffusive regime was  only briefly
considered in Refs.~\cite{minkovaboutptype,ColeridgePSiGe,minkovSO}. For the most ubiquitous 2D system known, 2DEG in Si-MOSFET, which fits well all theory requirements,  no convincing measurements of the EEC have been done so far. At the same time this system demonstrates positive magnetoresistance in parallel field, the behavior expected for EEC. In the 1980-s there were attempts to reveal EEC in Si from temperature and magnetic field dependences~\cite{bishop,davies,burdis}
of resistivity; these attempts were based on not yet developed theoretical concepts and did not lead to a self-consistent  picture of magnetotransport.

 Interest
to Zeeman splitting effects was resumed in 1997 with observation of a  huge
rise in resistivity of  2DEG  in clean Si-MOSFETs in parallel magnetic field~\cite{simonian, pudalovbpar, kravreview}, close to metal-to-insulator transition. The interest further increased with interpretation of this magnetoresistance as a signature of magnetic quantum phase transition \cite{QPT1,QPT2}.
In the 2000s several attempts to treat the parallel field magnetoresistance (MR) in terms of renormalization-group approach were taken both theoretically \cite{burmistrovshelk} and experimentally \cite{knyazev2006,kravnature}. This approach is in fact self-consistent generalization of the EEC for arbitrary interaction strength and conduction.
Independently, another
theoretical approach was developed in Refs.~\cite{GD,Das1,Das2}, and successfully applied~\cite{Dolgophighfield}, which accounts for resistivity increase with field  simply by  renormalization of the  density of states and single impurity scattering time. The latter effect is essentially different from logarithmic EEC  which emerges from multiple electron-impurity scattering.

The experimental situation, however, is more complicated:  studies~\cite{pudalov_0103087,Pud-prl_2002} showed a strong effect of disorder on the parallel field magnetoresistance, that was discussed in terms of the band tail effects in Refs. \cite{vitkalov_disorder,dolgopolov_disorder}.
 Moreover, detailed studies of the MR on different material systems~\cite{burdis,ColeridgePSiGe,vitkalovMR,  pudalovballistics, Klimov} did demonstrate quantitative disagreement between the fitted-to-EEC theory temperature- and magnetic field  dependences of the conductivity.

To summarize the present state of the field, there seems to be a common  agreement on the Zeeman nature of parallel field MR in  2D carrier systems.
However, two conceptually different underlining mechanisms of  MR  were put forward:
(i) EEC (multiple-scattering effect) and (ii) screening change in magnetic field (single-scattering effect). Which of them
is responsible for the experimentally observed strong MR in parallel field? The answer is especially crucial in the vicinity of  the metal-to-insulator transition, where the MR is dramatically strong.
Unfortunately, both theories become inapplicable in this regime of small conductances $\sigma \sim e^2/2\pi\hbar$. To address this issue, we have chosen to approach the problem from the large  conductance regime, where  both theories have solid ground, though  the MR is low.

In our paper we contest possible origin of the parallel field magnetoresistance of weakly interacting 2D electron gas.
In order to study EEC, we  take detailed measurements of the magnetoresistance tensor  in tilted field, and analyze the data  using the procedure developed in Refs.~\cite{minkoveefirst,minkovaboutptype}.   We stress that our approach does not rely on any particular microscopic theory, rather, it is {\em ab-initio} phenomenological and uses only general
property of the EEC in the diffusive  regime  to affect $\sigma_{xx}$ solely.

For the experiments we have chosen the simplest model system, the 2D electron gas in Si in diffusive regime $k_BT\tau/\hbar\ll1$, $\sigma\gg e^2/2\pi\hbar$. To vary the strength of the Zeeman splitting, and thus, the EEC magnitude, we  tilted magnetic field with respect to the 2D plane. This procedure allowed us to extract EEC on top of other magnetoresistivity effects and to establish two principally different regions: (i) high-field region,  where EEC  depends on total field and quantitatively agrees with the theoretically predicted $\ln(B)$ asymptotics, and (ii) low-field region,
where  EEC unexpectedly depends on perpendicular field component, grows with field and does not match existing theories.

 Our observations suggest a new insight on the origin of the parallel field MR: (i) the high-field small and $T-$independent EEC $\Delta\sigma(B)\propto\ln(B)$, revealed in the current study,   can not be responsible for large and $T$-dependent parallel field MR, and (ii) application of even low $g\mu_BB\sim k_BT$ perpendicular field component  strongly suppresses MR. The latter fact proves not purely spin nature of the parallel field MR and points at the incompleteness of the existing theory of MR.

The paper is organized as follows: after a brief description of  experimental details in  Section II, we give theoretical background  in section III, and describe the results  in Section IV, first on the EEC, then on the experimental proof of the Zeeman origin of the magnetic field effect on EEC, and further we compare the known EEC with the parallel field magnetoresistance to show that the EEC
is not the main origin of the  parallel field magnetoresistance.
 In Section V we discuss the obtained result and suggest possible directions of further development.

\section{EXPERIMENTAL DETAILS}
 The AC-measurements (13 to 73\,Hz) of the resistivity were performed
at temperatures 0.3-25\,K in magnetic fields up to 15 T
with two (100) Si-MOS samples with 200 nm oxide thickness:
Si-40, (peak mobility $\mu _{\rm peak} = 0.2~$m$^2$/Vs at $T = 0.4$~K), Si-24 (0.22m$^{2}$/Vs).
The samples were lithographically defined as rectangular Hall bars of the size 0.8$\times$5~mm$^2$.
To obtain the data for different orientations of magnetic field  relative to the 2D plane, the sample platform was rotated
{\it in situ} at low temperature using a step motor. Because of the smallness of the studied effects and unavoidable misalignment of potential contacts there always was some asymmetry (within less than a few percents) of both
$V_x$ and $V_y$ with respect to field reversal at a constant $j_x$  ($\sim 50$ nA) current direction. To compensate this asymmetry the field was swept from positive to negative values and the data were symmetrized. The alignment of the sample parallel to magnetic
field was done using the resistivity peak due to weak localization. The carrier density $n$
was varied by the gate voltage in the range $(8-35)\times 10^{11}$~cm$^{-2}$. The field was tilted in the $zy$-plane, always perpendicular to the current direction, which is however not crucial for Si because of weak spin-orbit  coupling~\cite{Pud-prl_2002}.

We have selected  samples with moderate carrier mobility  in order to ensure studies deep in the diffusive regime
$0.005<k_BT\tau/\hbar<0.2$, where EEC theory should be applicable, and at the same time
to achieve $\mu B\sim 1$ in available magnetic field. Other features which make Si-MOS system preferable for
this study are (i) short-range and uncorrelated scatterers which make motion diffusive, (ii) large conductance $k_Fl\sim20- 50$, which ensure quantum correction theory applicable,  (iii) very thin potential well  $<5$\,nm which excludes
orbital effects in parallel magnetic fields up to 20 T, and (iv)
filling of the lowest size quantization subband solely for $n<3.5\times10^{12}$\,cm$^{-2}$ with effective mass $m^*\approx0.2m_e$~\cite{ando}.

\section{THEORETICAL BACKGROUND}
The idea of the present
study is based on a property of the e-e correction to affect only
diagonal component of the conductivity tensor \cite{altshulerhall,Houghton}:
\begin{equation}
 \sigma=\frac{ne\mu}{1+\mu^2 B_{\perp}^2} \left( \begin{array}{cc}
 1  & \mu B_{\perp} \\
-\mu B_{\perp}  & 1 \\
\end{array} \right)
+ \left( \begin{array}{cc}
 \Delta\sigma    & 0 \\
0   & \Delta\sigma \\
\end{array} \right).
\label{aamctensor}
\end{equation}

Correspondingly, the procedure of the EEC extraction~\cite{minkoveefirst} for arbitrary  ($\Delta\sigma/ne\mu$) value  includes the following steps:
(i) Reversing the  measured resistivity tensor one calculates the conductivity tensor.
(ii) From $\sigma_{xy}$  one finds the mobility $\mu$ value, using  experimentally determined  density  $n$ \cite{commentondensity}.
(iii) Subtracting $ne\mu /(1+\mu^2 B_{\perp}^2)$ from $\sigma_{xx}$ one finds the
correction $\Delta\sigma(B,T)$.
The mobility $\mu(B,T,n)$ value is a by-product of this algorithm.

If the conductivity is large, $ne\mu \gg\Delta\sigma\sim e^2/2\pi^2\hbar$, the above algorithm leads to the following expressions for the  resistivity components:
\begin{equation}
\rho_{xx}=1/ne\mu\times[1-(1-\mu^2B^2)\Delta\sigma/ne\mu]
\label{aamrtensor}
\end{equation}
\begin{equation}
\rho_{xy}=B_\perp/ne\times(1-2\Delta\sigma/ne\mu)
\label{simpleDs}
\end{equation}
As we checked,  the{\b se formulae are} valid for all our data with $\rho_{xx}<4$ kOhm.

In practice, both $\Delta\sigma$ and mobility $\mu$ in Eqs. (\ref{aamctensor},\ref{aamrtensor},\ref{simpleDs}) might depend on $B$ and $T$ thus incorporating other possible magnetoresistance effects. Nevertheless, as follows from Eq.(\ref{simpleDs}), no matter how the mobility depends on field, the correction to the Hall coefficient arises from EEC solely.

Depending on the geometry of
experiment the following cases are possible: (i) The field is directed  normally
to the sample plane, $B=B_{\perp}$, during the sweep
(this case refers to most of the previous studies).
(ii) The field is inclined by angle $\theta$  relative to the 2D plane,
in order to enhance Zeeman effects which depend on
$B=B_{\perp}/\cos (\theta)$. (iii) The field  magnitude $|B|$
remains constant while sample is rotated relative to the field direction; in the latter case one expects the Zeeman effects to remain unchanged, thus leading to the angle-independent $\Delta\sigma$.
We note here that in the case of geometry (ii) data processing requires knowledge of the ratio $B_\perp/ne$  solely, rather
than the tilt angle. According to our definition of density~\cite{commentondensity}, the $B_\perp/ne$  ratio is obtained  from  the measured linear-in-total-field high temperature limit of the Hall resistance $R_{xy}^{HT}(B)=B_\perp/ne$ . We find this limit by linear extrapolation of $R_{xy}(B)$ to the highest temperatures (typically 15\,K), and by ignoring low-field effects  (see below).

Making use of geometries (ii) and (iii) is the key feature of the present study. We note here that
the $g$-factor in Si is large ($g=2$) and isotropic, which insures the Zeeman effects to be tilt independent. In systems with anisotropic $g$-factor,
like holes in GaAs~\cite{minkovaboutptype} one should take into account different components of the $g$-factor tensor, which complicates the problem.

According to the theory of interaction corrections~\cite{altshuleraronovlee,FinJETP} in its modern form \cite{ZNALarge,punnoose,burmistrovshelk},
the EEC value at zero magnetic field in the diffusive regime is \cite{commentonlnTEF}.
\begin{equation}
\Delta\sigma=\frac{e^2\ln (k_BT\tau/\hbar)}{\pi^2\hbar}\left(1+n_T \left\{  1-\frac{\ln(F_0^\sigma+1)}{F_0^\sigma} \right\}\right),
\label{ZNAmaineq}
\end{equation}
where $F_0^\sigma$ is a Fermi-liquid constant, $n_T=4n_v^2-1$ is the number of triplet channels of interaction, $n_v$- valley degeneracy (in the original formula  \cite{altshuleraronovlee,FinJETP} $n_T=3$).  In (001) Si-MOSFETs, electron system is 2-fold valley degenerate,  and the  degeneracy, if perfect, should increase $n_T$ to 15~\cite{finkelsteintriplets}. In fact, however, this degeneracy is never perfect, because of the
two sample-dependent parameters, a finite valley splitting $\Delta_v$ ~\cite{ando},
and intervalley scattering time $\tau_v\sim 10\tau\approx 1/4$~K$^{-1}$ for the sample Si-40~\cite{kuntsevich}. Both effects decrease the number of triplet terms, which was described theoretically~\cite{burmistrovshelk} and  studied
experimentally~\cite{Klimov,kravvv}.

The problem with MOSFETs is that $\Delta_v$ (which is typically less than $1/\tau$) can hardly be measured directly in zero and low magnetic fields, because valley splitting of Shubnikov-de Haas oscillations
does not exceed level broadening and hence can not be resolved \cite{anomalousDv}.  One could treat $\Delta_v$ as an additional free parameter, which vary in the range from 0 to $1/\tau$, strongly affecting predictions of the e-e interaction correction theory \cite{punnoose,burmistrovshelk}. Within the present study, uncertain $\Delta_v$ value affects only the effective number of valleys $n_v$
that can vary from $n_v=2$ to $n_v=1$. We therefore can use $n_v$ as adjustable parameter, which quantifies effective degree of valley multiplicity.

In 2001, the EEC was recalculated by Zala et al \cite{ZNALarge} and a new ``ballistic'' contribution was introduced, that gave explanation to $\rho(T)\propto T$ dependence observed in different 2D systems in the regime $k_BT\tau/\hbar>1$ \cite{proskuryakov, pudalovballistics}. It was shown experimentally \cite{liballistictodiffusion,renardballistictodiffusion,minkovballistictodiffusion} that  the ballistic and diffusive corrections differ fundamentally: diffusive EEC does not affect Hall component of conductivity tensor (see Eq. \ref{aamctensor}), whereas ballistic contribution is basically renormalization of the single impurity scattering time or mobility.
In the present study we do not consider ballistic contribution, rather, we extract from the experimental data and analyse the diffusive part of the EEC solely.

Theoretical prediction for the Zeeman splitting dependence of the EEC was  first given in \cite{LeeRamakr}:
\begin{equation}
\Delta\sigma=-\frac{e^2}{2\pi^2\hbar}\lambda_Dg_2(h),
\label{LeeRamakr0}
\end{equation}
where $h=g\mu_BB/k_BT$, and the two asymptotics for $g_2(h)$ function are:
\begin{equation}
g_2(h)=0.084h^2, h\ll1
\label{LeeRamakr1}
\end{equation}
\begin{equation}
g_2(h)=\ln(h/1.3), h\gg1
\label{LeeRamakr2}
\end{equation}

Within the same theoretical formalism, the Zeeman splitting effect on ballistic and diffusive corrections to magnetoresistance in parallel field was recalculated in Ref.~\cite{ZNABPar}. Although the theory was successfully used to fit some data on parallel field magnetoresistance~\cite{vitkalovMR}, the procedure of the comparison with theory is ill-defined and requires careful separation of the diffusive and ballistic EEC contributions in the crossover regime.
Indeed, theoretical EEC is a sum of two contributions with absolutely different structure: (i) a ballistic one which comes from the renormalization of single impurity scattering and (ii) diffusive EEC for which $\Delta\sigma_{xy}=0$. Since our method catches only diffusive part,
we briefly discuss below modern theoretical expressions in the diffusive regime $k_BT\tau/\hbar\ll1$ solely
\begin{equation}
\Delta\sigma(h)=\frac{e^2}{2\pi^2\hbar}n_v^2\frac{0.091F_0^\sigma}{(1+F_0^\sigma)^2}h^2 , h\ll1
\label{ZNABpar0}
\end{equation}
This low field limit is close to \cite{LeeRamakr} for small $|F_0^\sigma|\ll 1$. In the high field limit, the diffusive contribution from Ref.~\cite{ZNABPar} is given by:
\begin{equation}
\Delta\sigma=\frac{e^2}{2\pi^2\hbar} 2n_v^2 \left( 1-\frac{\ln(F_0^\sigma+1)}{F_0^\sigma} \right) \ln \frac{h}{F_0^\sigma+1}, h\gg1
\label{ZNABpar1}
\end{equation}

Noteworthy, functional dependence on $F_0^\sigma$ is the same for Eqs.~(\ref{ZNAmaineq}) and (\ref{ZNABpar1}). This fact has a transparent physical meaning: application of high field $h\gg 1$ suppresses temperature dependence of only $2n_v^2$ triplets with $s_z=0$. This suppresion comes from expansion of the $\ln(h)=\ln (g\mu_BB)-\ln (k_BT)$. Correspondingly, if we define
\begin{equation}
\lambda=( 1-\ln(F_0^\sigma+1)/F_0^\sigma ),
\label{eq:F0s}
\end{equation}

 we may rewrite the theoretical expectation for the low temperature ($T\ll1/\tau\ll E_F$) high field ($k_BT\ll g\mu_BB\ll E_F$) asymptotics:
\begin{equation}
\Delta\sigma(T,B)=C+\frac{e^2}{2\pi^2\hbar}\left[(1+2n_v^2\lambda-\lambda)\ln (T) +2n_v^2\lambda\ln(B)\right]
\label{lowTBasympt}
\end{equation}
Here $C$ is a $T-$ and $B-$ independent term.  This expression allows one to compare experimental data on low-temperature high-field asymptotics of the EEC with the microscopic theory predictions. It's meaning is as follows: in the high field limit $h\gg 1$, the magnetic field dependence is not affected by temperature. In Appendix A we show that from the practical point of view this limit is achieved  already for $h>2$.

To conclude this section, studying magnetoresistance in purely parallel field is insufficient to disentangle the ballistic and EEC contributions. Tilting the field  enables one to overcome this drawback.
As  shown in the next section, application of  high tilted fields allows us
to achieve in  experiment  the asymptotical $\Delta\sigma(T,B)$ behavior, for which there is a firm theory prediction.

\section{Results}
This section is organized as follows: in the first subsection we discuss phenomenology, low-field regime and experimental proof of the EEC dependence on modulus of magnetic field rather than on its direction. In the second subsection we consider high-field asymptotics of the EEC, and compare them with the Fermi-liquid theory expectations. In the third subsection we discuss magnetoresistance in parallel and tilted fields.

\subsection{Tilt angle independence of the EEC}
\begin{figure}
\begin{center}
\centerline{\psfig{figure=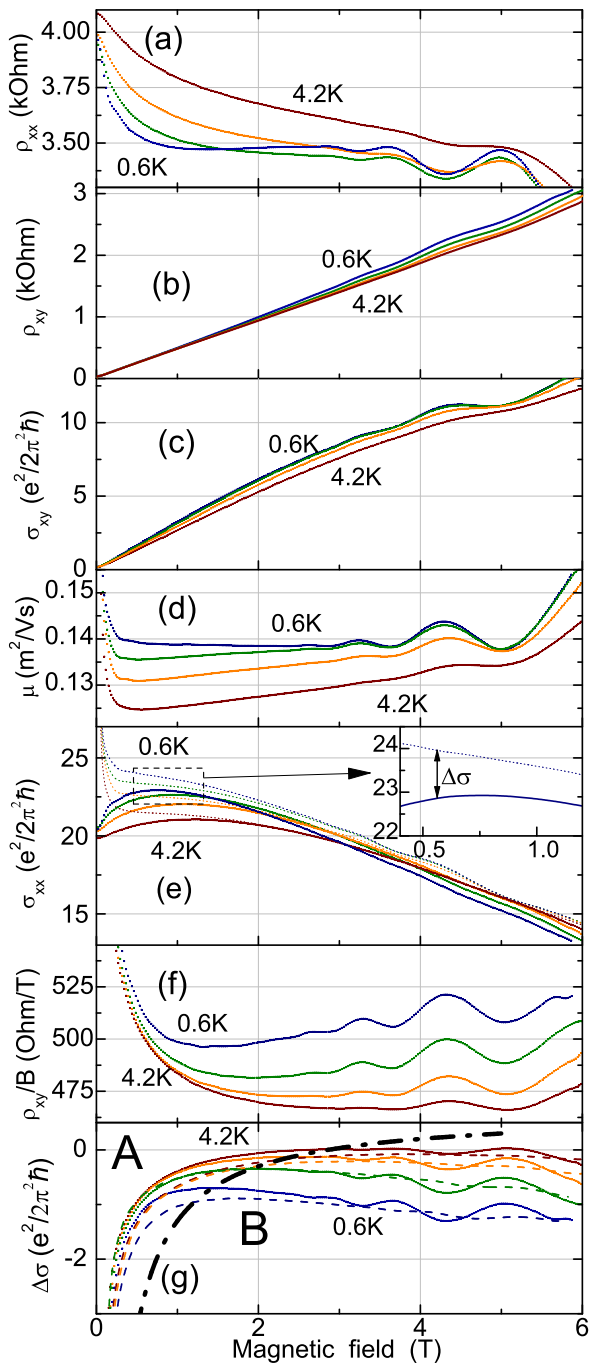, width=230pt}}
\caption{(Color online)Magnetoresistivity tensor components for various
temperatures (0.6K, 1.3K, 2.5K, and 4.2K):  (a) $\rho_{xx}$, and  (b) $\rho_{xy}$  versus perpendicular magnetic field.   (c) Hall conductivity $\sigma_{xy}$  calculated from the data.  (d) Mobility $\mu$, calculated from the Hall conductivity using Eq. (\ref{aamctensor}). (e) Solid lines: $\sigma_{xx}$ calculated by  inverting the measured resistivity tensor (panels (a) and (b)). Dotted lines: Drude value of $\sigma_{xx}$ calculated using mobility from panel (d). The  inset
blows up the data to show graphically the definition of $\Delta\sigma$.  (f) Hall angle $\rho_{xy}/B$ for the same data set.  (g) Solid lines: EEC versus magnetic field recalculated from the data. Dashed lines: EEC at the very same set of temperatures versus total magnetic field for the sample tilted by
$45^\circ$. Bold line separates the  low-field and high-field regimes, denoted  A  and B,  respectively. Sample Si-40,  electron density $n=1.25 \times 10^{12}$cm$^{-2}$.}
\label{Perpfield}
\end{center}
\end{figure}

We explore  MR in the range of fields $B_{\rm tr}< B_\perp<1/\mu$, i.e. in the domain between weak localization, $B_{\rm tr}= (hc/e) (1/4 \pi l^2$), and Shubnikov-de Haas oscillations (here $l$ is the transport mean free path). Figure \ref{Perpfield}a shows  magnetoresistance $\rho_{xx}$ and Hall resistance $\rho_{xy}$ versus $B_{\perp}$ in  field up to $1/\mu\approx 6$\,T.  The data were collected at various temperatures (0.6-4.2K) in a standard geometry (magnetic field perpendicular to the sample plane) for sample Si-40 at  $n=1.25\times 10^{12}$cm$^{-2}$. The set of curves for Si [Fig.~\ref{Perpfield}(a)] does not look similar to numerous data for n-GaAs 2D systems~\cite{paalanen,taboryski,choi,poirier,minkoveefirst,liballistictodiffusion,renardballistictodiffusion}. Indeed, there are
two important features:
(i)unlike GaAs, for Si in high fields ($>1$\,T in Fig.~1(a))
the MR gets weaker and ultimately even changes sign to positive as temperature decreases;
(ii)  the Shubnikov-de Haas oscillations start to appear already in  fields $B_{\perp}<\mu^{-1}\approx 6T$,  due to short-range disorder.

Because of the above features, the  EEC  is not seen straightforwardly  from the data. Therefore, in order to extract it we use the procedure described in the previous section.
At the first step we invert the resistivity tensor and obtain the conductivity one (solid lines in Figs. \ref{Perpfield}c, \ref{Perpfield}e).
Hall component of the conductivity tensor allows one to calculate mobility $\mu$ using Eq.~(\ref{aamctensor}) (shown in Fig. \ref{Perpfield}d).
Surprisingly, the mobility appears to be field-dependent (unlike that  in experiments
with GaAs \cite{minkoveefirst}): in strong fields ($B>1$\,T),
the higher is the temperature the stronger is the field dependence.
 This effect is beyond the scope of the present study,  we note only that the sign of the $\mu(B)$-dependence is in-line with expectations from  memory effects~\cite{memory} though such effects do not produce temperature dependence.  Field dependent mobility was suggested to arise also from ballistic correction $\delta\sigma_{xy}(B,T)\propto\sqrt{T}/B$ \cite{gornyimirlin}. Even though such correction looks  qualitatively similar to our data, it is hardly relevant, because originates from multiple cyclotron returns and should be valid for $k_BT\tau/\hbar\gg1, \hbar\omega_c\gg k_BT$, which is not the case.

In weak fields, the  Hall mobility  increases as $B$ decreases due to increase of the Hall slope. Having the mobility known, we calculate the Drude part of $\sigma_{xx}$,   $ne\mu/(1+\mu^2B^2)$ [shown by dotted lines in Fig.~\ref{Perpfield}(e)]. Correction to conductivity $\Delta\sigma$ is calculated as a difference between $\sigma_{xx}$ and its Drude expectation (see inset to Fig.~\ref{Perpfield}e  for the graphical definition).   The resultant $\Delta\sigma(B)$ is shown by solid lines in Fig.~\ref{Perpfield}f. As it is clear from Eq.~(\ref{simpleDs}), the correction resembles behavior of the Hall coefficient $\rho_{xy}/B$ (shown in Fig. \ref{Perpfield}f): the larger the Hall coefficient, the less the correction.

All  $\Delta\sigma(B_{\perp})$ dependences (collected at different temperatures, densities, tilt angles and for different samples) manifest  similar behavior: in low fields (region A in Fig.~\ref{Perpfield}g) $\Delta\sigma$ grows as $B$ increases, then reaches a maximum and decreases in  the high field region B.
In the low field region A, the  feature in $\Delta\sigma$  originates from non-linearity of the Hall resistance with field  (as  seen from Fig.~\ref{Perpfield}f). Similar low-field feature was observed in numerous previous studies with various 2D systems \cite{HallZhang,HallNewson,Goh,MinkovHall}; it is still poorly understood. In Appendix B we summarize our observations on the low-field Hall nonlinearity and argue that this low-field feature does not follow contemporary theories.  Empirically, the boundary between the regions A and B (a point where $\Delta\sigma$ is maximal) roughly follows the equation $B_{\rm crossover}\approx\sqrt{B_{tr}^2+(k_BT/g\mu_B)^2}$.

The correction decreasing with increasing field in the region B Fig.~\ref{Perpfield}g is qualitatively in line with Zeeman splitting effect \cite{LeeRamakr,ZNABPar,minkovaboutptype}. The latter should be direction independent, as discussed above. To check this property we set the tilt angle to 45$^\circ$, measured both $\rho_{xx}$ and $\rho_{xy}$ for the same temperatures and recalculated the EEC.
The resultant EEC versus {\em total} magnetic field is shown in Fig.~\ref{Perpfield}g by dashed lines. In the high-field region B, the monotonic parts of the EEC data (ignoring Shubnikov-de Haas oscillations)  in perpendicular and tilted field are quantitatively similar to each other,   even though the perpendicular component of the field differs by a factor of $\sqrt{2}$. This observation proves the Zeeman nature of the EEC in the high-field region.

\begin{figure}[t]
\begin{center}
\centerline{\psfig{figure=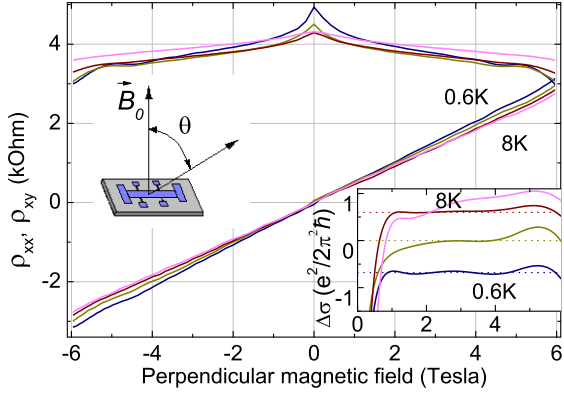, width=220pt}}
 \caption{(Color online) Resistivity tensor components  for sample Si-40 at different temperatures (0.6, 1.7, 4, 8K) versus perpendicular to the sample plane field component $B_{\perp}=B_0\cos(\theta)$. $n=1.25\times10^{12}$~cm$^{-2}$. The data were obtained by sweeping angle $\theta$ (geometry of the experiment is shown in the left part of the figure) at constant magnetic field $B_0=6$T. Inset shows EEC versus $B_{\perp}$  recalculated from the data in the main panel (solid curves). Dotted lines - the corresponding field-independent values of the EEC. }
\label{Spining}
\end{center}
\end{figure}

Another instructive way to check whether
$\Delta\sigma$ depends on the {\em total} field is to rotate the sample in constant field.
During these measurements, the tilt angle $\theta$ was slowly swept at a constant rate using a step motor. We used positions of the specific sharp mirror symmetric  features  in the $\rho_{xx}(\theta)$ dependence at $\theta=0$,~$\pi/2$,~$\pi$ to  calibrate angle and calculate perpendicular component of the field. Figure \ref{Spining} shows $\rho_{xx}(B_{\perp})$ and $\rho_{xy}(B_{\perp})$ for the same electron density and different temperatures.
The inset to Fig. \ref{Spining} shows the corresponding EEC. It is easy to see that gradual changes of $\Delta\sigma$ with $B_\perp$, which were seen in Fig.~\ref{Perpfield}g almost disappeared in the inset to Fig.~2; this demonstrates that EEC remains unchanged in constant total field.

We stress the importance of this ``rotating field'' experiment; indeed, the coincidence  of $\Delta\sigma(B)$ curves in Fig. 1f for different tilt angles is not a complete proof for direction-independence of the EEC,  because it does not
exclude corrections to conductivity $\propto\ln(B_\perp)$ (see e.g. Ref.\cite{AALK}). In the latter case one would see just a constant shift $\propto\ln(cos(\theta))$ in tilted field sweep experiment (Fig. 1f) with no visible change in functional form of the $\Delta\sigma (B)$ dependence. In contrast, in the rotating field experiment such corrections would reveal themselves as
inclined $\Delta\sigma(B_\perp)$-lines, which is not the case  in the insert to Fig. 2.

\subsection{High-field asymptotics of the EEC}
The
 discussed above field  direction independence of the EEC was tested with two samples in the density range $n= (1 -3) \times10^{12}$ cm$^{-2}$ which corresponds to $\sigma=(2-20)$ $ e^2/(2\pi\hbar)$, and in the field range $B_\perp <1/\mu$.

The data in the high field region B ($B>T$) should follow the asymptotics of Eq.(\ref{lowTBasympt}).
The field range for observing the logarithmic-in-field behavior, however,  is limited on the low-field side  by
low-field Hall feature (at $B_\perp\sim1$\,T) and, on the high field side, by the onset of Shubnikov-de Haas oscillations (at  $B_\perp\sim 1/\mu\sim 6$ T). For this reason, the field range for fitting the data with $\ln (B)$-dependence is  less than one decade.

We  performed detailed measurements of magnetoresistivity tensor at different tilt angles in magnetic fields up to 15 T.
 The high-field asymptotes were fitted with a function
\begin{equation}
\Delta\sigma(B)=e^2/(2\pi^2\hbar)\times(-D\ln (B)+A\ln(T)+C),
\label{10}
\end{equation}

 where $A$ and $D$ are two positive adjustable parameters, common  for all curves.
According to Eq.(\ref{lowTBasympt}):
$A= (1+2n_v^2\lambda-\lambda)$ and $D=-2n_v^2\lambda$.

\begin{figure}[t]
\begin{center}
\centerline{\psfig{figure=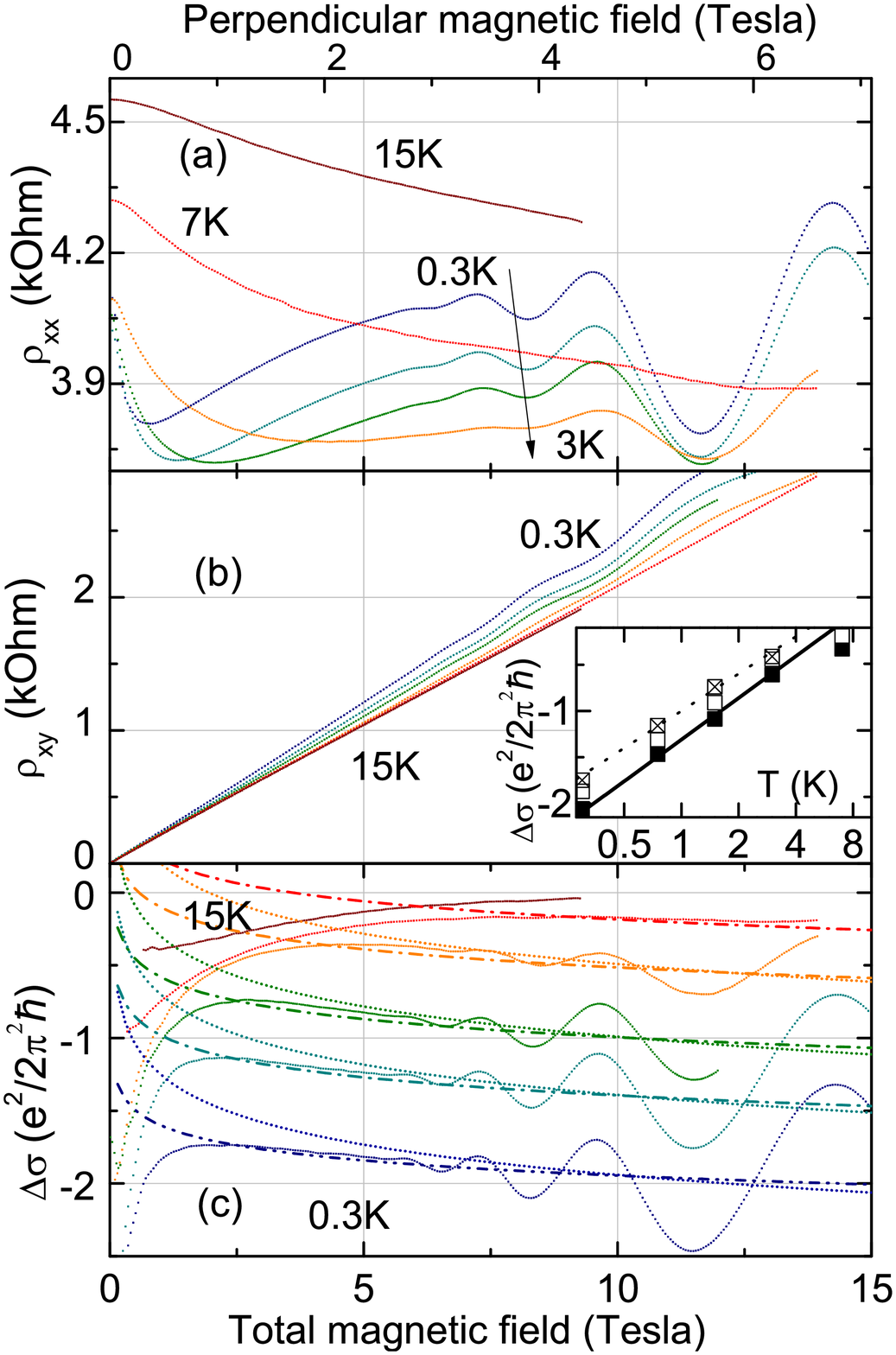, width=220pt}}
 \caption{(Color online) a),b)Resistivity tensor components  for sample Si-40 at various temperatures (0.3, 0.7, 1.6, 3, 7, and 15K) versus total magnetic field (bottom axis) and perpendicular component (top axis). $n=1.25\times10^{12}$~cm$^{-2}$; $\theta=62^\circ$. Panel c) shows EEC  calculated from the data in the panels a) and b) (solid curves). Dash-dotted lines -- set of functions $\Delta\sigma=-0.15\ln (B)+C$; dotted lines $\Delta\sigma=-0.3\ln (B)+C$ ($C$-values are adjusted to fit the high-field asymptotes of the EEC). An inset to panel (b) shows the obtained $T$- dependence of the EEC  at $B=3$\,T (crossed rectangles), 7.5\,T (empty rectangles) and 15\,T (filled rectangles).  Solid line - linear-in-$\ln(T)$ fit of $\Delta\sigma(T,15\rm T)=\frac{e^2}{2\pi^2\hbar}\times0.65\ln(T/7.5$ K$)$, dotted line - linear-in-$\ln(T)$ fit of $\Delta\sigma(T,3\rm T)=\frac{e^2}{2\pi^2\hbar}\times0.58\ln(T/5.6$ K$)$.}
\label{15tesla}
\end{center}
\end{figure}

\begin{figure}[t]
\begin{center}
\centerline{\psfig{figure=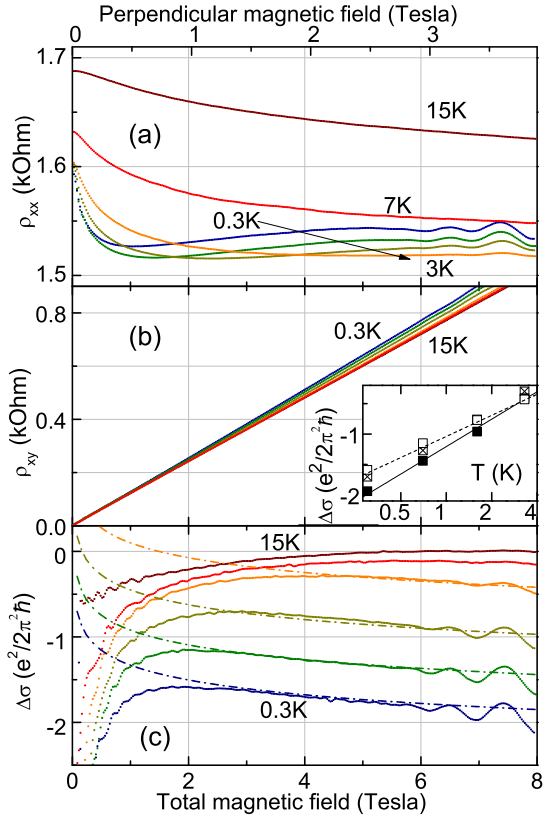, width=220pt}}
 \caption{(Color online) a),b)Resistivity tensor components  for sample Si-40 at
various temperatures (0.3, 0.7, 1.6, 3.3, 7, and 15\,K) versus total magnetic field (bottom axis) and perpendicular component (top axis). $n=2\times10^{12}$~cm$^{-2}$; $\theta=62^\circ$. Panel c) shows EEC  calculated from the data in the panels a) and b) (solid curves). Dash-dotted lines -- set of functions $\Delta\sigma=-0.22\ln (B)+C$ with different $C$, which gives a satisfactory high-field asymptotes of the EEC (see text). An inset shows the obtained $\Delta\sigma(T)$ dependence collected at 2~T (empty rectangles), 4~T(crossed rectangles) and 8~T (black rectangles). Solid line is a fit of the high-field (8T) data with logarithmic function $\Delta\sigma=\frac{e^2}{2\pi^2\hbar}\times0.65\ln(T/6.4$ K$)$; dashed line is a fit of the low-field (2T) data with logarithmic function $\Delta\sigma=\frac{e^2}{2\pi^2\hbar}\times0.5\ln(T/8$ K$)$}
\label{Vg21}
\end{center}
\end{figure}

Example of the data and corresponding EEC is shown in Fig.~\ref{15tesla}. One can see that the $\Delta\sigma(B)$ dependencies for different temperatures follow almost parallel to each other being only shifted vertically; exactly such  behavior should be expected for the EEC correction according to Eq.(\ref{10}).
The temperature prefactor  $A$ can be easily found from the corresponding $\Delta\sigma(T)$- dependence (see the inset to Fig.~\ref{15tesla}b); for the given dataset $A=0.6\pm0.04$.
The data in the  inset also demonstrates that the obtained  $A$ value is field-independent  for $B>5$\,T, which confirms that the analyzed data follow the high-field asymptotic behavior. As for the field-dependence prefactor $D$, since the accessible field interval is less than a decade and we don't know how far the low-field correction (caused by the Hall anomaly) may extend, we can only roughly estimate  $-0.3<D<-0.15$. Its lower bound is found from the low-field,  $B<6$\,T, data and the upper bound - from the high field ($B>10$~T) data.

The next logical step would be to analyze the density dependences of the two prefactors, $A$ and $D$, and to check  their consistency with  each other and with  a microscopic theory. 
Figure \ref{Vg21} shows magnetoresistance and the corresponding EEC for elevated density $n=2\times 10^{12}$~cm$^{-2}$  and for the same tilt angle. Since conductivity  increases here by a factor of 2.5, the effect of EEC becomes less pronounced  on  top of Drude conductivity, nevertheless both temperature and magnetic field  dependences of the EEC remain approximately the same: $A=0.65$; $D=0.2- 0.4$.

The resulting  $A$ and $D$-values are summarized in Table~\ref{F0summarytable}. As we argued above, these parameters do not demonstrate a pronounced density dependence. This fact is reasonable because in the studied range of densities (i) interaction parameter $r_s=(\pi n)^{-1/2}/a^*_B$ \cite{ando} is rather small  $\sim 1.5  - 2.8$ , and (ii) conductivity is large compared to the quantum unit to ensure smallness of the renormalization effects\cite{finkelsteintriplets, minkovdecrconductance}.  In Table~\ref{F0summarytable} we present $F_0^\sigma$ calculated from  $A$ and $D$- values using Eqs.~(\ref{eq:F0s}),(\ref{lowTBasympt})  for the effective valley multiplicity $n_v=1$ and  2.
 For comparison, we also show the $F_0^\sigma$-values experimentally determined from  Shubnikov-de Haas oscillations \cite{PudFL} and from $R(T)$ ballistic dependences \cite{Klimov}.
Firstly,  $|F_0^\sigma|$ extracted from $A$ decreases with density.
Secondly,  $|F_0^\sigma|$ values extracted from $A$ for $n_v=2$ is rather close to the earlier measured values. If we adopt the effective valley multiplicity $n_v$ to be somewhat smaller than 2, the agreement will become even better. When making such comparison with earlier data one should keep in mind that previous  results were obtained  in high-temperature ballistic regime, where both valleys contribute equally and $n_v=2$ exactly. One should also take into account that the $|F_0^\sigma|$ values for the diffusive (this work) and ballistic (previous data \cite{Klimov,PudFL}) regimes should not coincide, for the reasons discussed in Ref.~\cite{ZNALarge}. We therefore conclude the temperature dependence of the diffusive EEC (i.e., $A$-values) to agree quantitatively with earlier data and to agree qualitatively  with theory.

Possible  reasons for large uncertainty  in $D$-values and their poor consistency with $A$-values may be (i) too narrow range of fields accessible  for identifying  $\ln (B)$ dependence (see above),  (ii) the effective number of valleys ($1<n_v<2$) may be different in the field and temperature dependences.
For the lowest studied density $n=0.9\times10^{12}$~cm$^{-2}$, the low-field feature broadens and obscures  observation of the decreasing logarithmic
$\Delta\sigma (B)$ dependence; this prevented us from measuring the respective  $D$ values.

To summarize this section, we presented above three firm experimental observations on the high-field and low-temperature behavior of the EEC, as follows: (i) the observed $\Delta\sigma(B,T)$ dependences do not demonstrate any anisotropy with respect to the field direction,
 (ii) $\Delta\sigma$ is linear both in $\ln(B)$ and $\ln(T)$ in line with the expected 
 asymptotics (Eq. \ref{lowTBasympt}), and (iii) the observed field and temperature dependences have an anticipated
 magnitude. In total, the above listed facts prove that the field and temperature dependences of conductivity indeed represent the EEC. The prefactor in the temperature dependence of $\Delta\sigma$ reproduces  $|F_0^\sigma|$ values that are decreasing with density and are reasonably consistent with earlier data. 
The prefactor in the $\ln(B)$-dependence  does not contradict the microscopic theory and earlier data though it was determined with rather large uncertainty.

\begin{table}
\caption{Summary of interaction parameters found from the EEC measurements.\label{F0summarytable}}
\begin{tabular}{|c|c|c|c|c|c}
  \hline
    $n$,$10^{12}$cm$^{-2}$& 0.9 & 1.25&  2&3&\\

  \hline
  $\rho_D$, kOhm & 8& 4 &2&0.9 \\
  \hline
  $r_s$ & 2.77& 2.35 &1.86&1.52 \\
  \hline
$F_0^\sigma$ exp. \cite{PudFL,Klimov}  &-0.25& -0.2 &-0.13&-0.076\\
 $A$ & $0.25$ &  0.6$\pm 0.04$&0.65$\pm 0.05$&$0.65\pm 0.06$ \\
 $F_0^\sigma (A)$ $n_v=1$&-0.72& $-0.52\pm0.04$  &$-0.47\pm0.05$&$-0.47\pm0.06$\\
  $F_0^\sigma (A)$ $n_v=2$&-0.2& $-0.12\pm 0.01$&$-0.1\pm0.02$&$-0.1\pm0.02$ \\
  $D$ & -- &  0.15 -- 0.3&0.2--0.4&0.2--0.4 \\
 $F_0^\sigma (D)$ $n_v=1$&--  & $-0.20\pm0.07$  &-0.17-- -0.32&-0.17-- -0.32 \\
 $F_0^\sigma(D)$ $n_v=2$&-- & $-0.06\pm0.02$ &-0.05-- -0.09&-0.05-- -0.09\\
  \hline
\end{tabular}
\end{table}

\subsection{Magnetoresistance}
A nonmonotonic magnetoresistance in perpendicular field and  at low temperatures in Si and p-SiGe has been  observed since 1982 \cite{PrusCooling,bishop,ColeridgePSiGe}: as field increases,  the  drop in $\rho_{xx}$ due to weak localization is followed by the resistivity increase (see, e.g. Fig.~\ref{Perpfield}a). Surprisingly, the effect has not been discussed and understood yet. As can be seen from our data  (Fig.~\ref{15tesla}a), field tilting makes this effect even more pronounced, thus pointing to the Zeeman nature of the nonmonotonic magnetoresistance. Qualitatively, this seems to be transparent: the larger is the Zeeman field, the stronger is the positive component of magnetoresistance. In the strictly parallel field orientation there is no weak localization suppression and the magnetoresistance is purely positive. Following Castellani et al.~\cite{Castellani}, the  magnetoresistance in the diffusive regime for 2D systems was often attributed to  EEC~\cite{knyazev2006} and used to evaluate interaction parameter~\cite{kravnature,proskuryakov}. However, whether the observed parallel field magnetoresistance may be entirely
attributed to EEC, has not  been proven so far.

\begin{figure}[t]
\begin{center}
\centerline{\psfig{figure=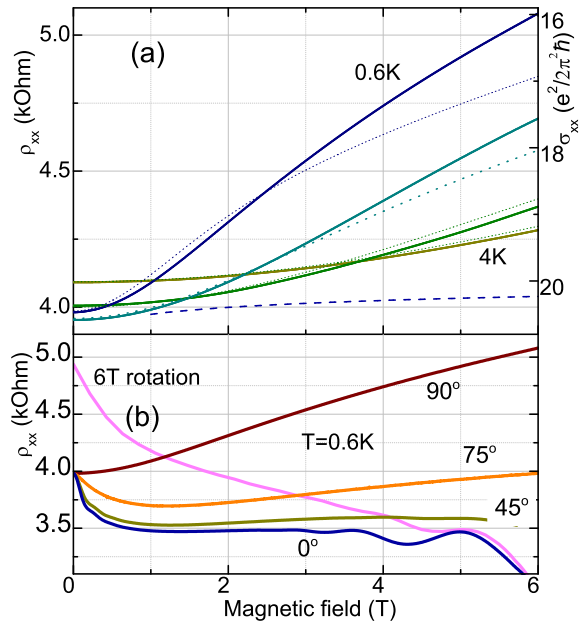, width=220pt}}
 \caption{(Color online) (a)Parallel magnetic field magnetoresistance for Si-40, $n=1.25\times 10^{12}$cm$^{-2}$ (solid lines). Direction of the temperature (0.6K,1.3K,2.2K,4K) change is indicated by an arrow. Right axis - conductivity units.
Dotted curves - theory prediction \protect{\cite{ZNABPar,vitkalovMR}} for 2-valley system, $F_0^\sigma=-0.2$, $\Delta_v=0$.
Dashed curve - experimentally determined high field limit of the EEC ($C+0.15\ln(B)$). 
(b) Resistivity versus total magnetic field  for Si-40, $n=1.25 10^{12}$cm$^{-2}$, $T=0.6$K for different tilt angles (shown in Figure) with the lowest curve being field perpendicular to the sample plane 0$^{\rm o}$, and the highest positive slope curve 90$^{\rm o}$ being parallel to the sample plane. ``6T rotation'' curve shows magnetoresistance versus perpendicular component of the field for fixed total field 6 T.}
\label{parallel}
\end{center}
\end{figure}

Having developed a technique  to measure Zeeman field dependence of the EEC directly, with no adjustable parameters, we can compare now  EEC with magnetoresistance. Figure \ref{parallel}a shows a set of magnetoresistance curves collected at different temperatures for $n=1.25\times 10^{12}$~cm$^{-2}$ in the parallel field orientation. High field asymptotics of EEC for the same sample and same density $\Delta\sigma=C+0.15\ln(B)$ is also shown in Fig.\,\ref{parallel}a by dashed line (the prefactor 0.15 for this particular density is found in the previous section). The observed magnetoresistance appears to be (i) an order of magnitude larger than EEC, and (ii) has much stronger temperature dependence; therefore, it  is mainly of a different origin. This is one of the key results of the present
study, it brings into a question the validity of usage parallel field magnetoresistance for finding the interaction constant.

We have shown above that the multiple scattering (EEC) approach leads to much smaller magnetoresistance than that observed; we test below whether or not single impurity scattering processes play dominant role in magnetoresistance.
We show in Fig.~\ref{parallel}a (dotted lines) the magnetocondutivity calculated according to Eq. (4)  from Ref. \cite{vitkalovMR} (two-valley version of theory \cite{ZNABPar}) for a system with $\Delta_v=0$, $\tau_v^{-1}=0$, and $F_0^\sigma=-0.2$. The calculated theory curve seems to demonstrate a reasonable agreement with the same experimental data. This agreement (in contrast to above sharp disagreement - see the dashed curve in Fig. \ref{parallel}a) has a simple physical explanation: the calculated curve comprises mainly (more than 65 \%) single impurity scattering renormalization, and only 35\% of the diffusive EEC.
We believe that by using two different $F_0^\sigma$ values for ballistic and diffusive contributions \cite{ZNALarge} and by introducing finite valley splitting and intervalley scattering rate \cite{punnoose} (i.e. about 4 adjustable parameters) one can achieve a perfect agreement with experiment. This exercise reproduces an apparent successful comparison of the parallel field magnetoresistance data in Si-MOSFETs\cite{Klimov,vitkalovMR} with theory\cite{ZNABPar}.

Is this the end of the story and complete understanding of the magnetoresistance in the 2D system? Our answer is NO; in order to argue this point we apply perpendicular field. Theoretically \cite{gornyimirlin} the Zeeman field-induced magnetoresistance should remain the same in tilted field. Figure~\ref{parallel}b shows magnetoresistance at  fixed $T=0.6$~K and $n=1.25\times 10^{12}$~cm$^{-2}$ for various tilt angles. In perpendicular field (0$^\circ$ curve), as field increases from zero,   weak localization becomes suppressed, leading to the negative magnetoresistance; further we observe almost flat magnetoresistance until the onset of Shubnikov-de Haas oscillations. Such classically flat magnetoresistance is a signature of a field-independent mobility.
In the opposite limit of parallel field (90$^\circ$ curve) we observe a huge, 25\% magnetoresistance in field of 6\,T. This fact itself is
remarkable because we did not expect much difference between the parallel and perpendicular configurations in fields $B>B_{\rm tr}$ within the above simple considerations.

Although we do not believe in orbital origin of the parallel field magnetoresistance, let us assume for a moment that only parallel component of the field affects mobility. If this is so, then  by
tilting the field from purely in-plane direction ($90^\circ$)
to slightly out of plane (75$^\circ$), when parallel component of the field decreases only by 4\%,
we would expect roughly the same  positive magnetoresistance as that in the purely parallel field. Surprisingly, after suppression of weak localization, the observed positive upturn in magnetoresistance is quite
shallow and is no more than $\sim$6\%! For intermediate tilt angle (45$^\circ$), the magnetoresistance  (Fig.~5\,b)  only slightly deviates from the perpendicular field data, which further confirms suppression of the parallel field MR by perpendicular component.

To summarize our observations, the positive magnetoresistance that is weak in perpendicular and tilted fields increases dramatically when the field turns  parallel to the 2D plane. Such behavior is counterintuitive and qualitatively different from that in GaAs-based structures.

\section{Discussion}
In the present study, by
using  magnetoresitance tensor measurements technique 
in  tilted magnetic field, we extracted  EEC and explored Zeeman effects in conductivity.
We demonstrated fruitfulness  of this approach and revealed the anticipated \cite{LeeRamakr} high-field logarithmic asymptotics of the EEC.

Our measurements also reveal  several puzzling features of the magnetotransport in  Si-MOSFETs.
Firstly, the low-field drop  of the EEC (i.e., the increase in Hall resistance), detected in our experiments, lacks an explanation. In order to clarify this issue one requires more precise very-low-field Hall measurements with
variety of samples ranging in mobility and conductance values, as well as  theoretical framework to treat the problem.

 Another important  observation is  that the parallel field magnetoresistance
is not produced by EEC solely. This observation poses a question of the applicability of the EEC theory to determination of the interaction constant from the parallel field magnetoresistance \cite{burdis,kravnature,vitkalovMR}. Indeed, our data prove that  there
is  another unexplored mechanismthat also contributes to parallel field magnetoresistance; it's necessary to  disentangle different MR mechanisms, before extracting the Fermi-liquid constant.

 Finally, our key result is that the parallel field magnetoresistance  is suppressed by rather insignificant nonquantizing perpendicular field component. This result directly contradicts to the predictions of the  electron-electron interaction theory \cite{gornyimirlin}.  Were the parallel field magnetoresistance
caused by EEC, such behavior could be straightforvardly explained by complete suppression of magnetoresistance already at $\mu B_{\perp}=1$, according to Eq.~(\ref{aamrtensor}).
However, we have shown above, that EEC is too small and alone can not be a reason for the parallel field magnetoresistance.
According  to Eq.~(\ref{aamrtensor}) it means that the parallel field MR comes from mobility renormalization.

 There are several potential mechanisms of such renormalization: (i)  mobility may drop with parallel field due to finite width effects \cite{Das2}. This mechanism, however,  is not supported by our data (compare the 90$^\circ$- and 75$^\circ$-curves in Fig.~\ref{parallel} that show a factor of 3 diminished magnetoresistance in almost the same parallel field), and is not very probable because of thin
potential well in Si-MOSFETs  ($\sim 3$nm) compared to  magnetic length $l_B$[nm]$=25.7(B_{||}$[T])$^{-0.5}$; (ii) mobility may depend on Zeeman splitting in agreement with the predictions of the screening theory \cite{GD, Das1,Das2}. The latter mechanism should produce direction-independent positive magnetoresistance, exactly the behavior observed in heavily-doped multisubband thin Si quantum wells\cite{Troup}.
This however does not reconcile with our data.
(iii) Surface roughness could also  affect MR through the weak localization suppression \cite{mensz}, however, this mechanism is hardly relevant because it produces a negative MR in contrast to our observations.

Experimentally, we observe a picture opposite to the common sense arguments: parallel field magnetoresistance is strongly suppressed when sample is rotated in fixed total field. This observation is not a property of  particular studied samples, rather,
there are numerous  examples in literature where strong positive MR in parallel field coexists with almost
shallow or even negative MR in perpendicular or tilted field: for p-GaAs~\cite{kumar}, strained Si~\cite{lai}, Si/SiGe quantum wells~\cite{okamoto}. Dramatic delocalizing effect of the perpendicular field was also reported for high mobility  Si-MOSFET samples in the low-density/high resistivity ($\rho \gtrsim 2\pi\hbar/e^2$) regime in the vicinity of the  metal-insulator transition~\cite{krav98}. Such behavior is not explained  at all, and  the question why parallel field MR is suppressed by the perpendicular field component remains open.

We believe that the parallel field magnetoresistance suppression in perpendicular field is somehow related to the low-field Hall feature (region A in Fig.~\ref{Perpfield}g). The empiric crossover between the low- and high-field regimes  $B_{\rm crossover}\approx\sqrt{B_{\rm tr}^2+(k_BT/g\mu_B)^2}$  points to the localization-related nature of the low-field Hall feature.
We, therefore, suggest the following model: a small part of electrons is localizedin low binding-energy states and coexist with
mobile electrons. We stress that these states are different from conventional tail of localized
states located below the mobility edge   \cite{dolgopolov_disorder,vitkalov_disorder}.
The localized electrons do not contribute to Hall effect at $B_\perp=0$ because they can be treated as zero mobility carriers.   
Let us assume also that the localized states promote strong parallel field MR. This assumption does not contradict to the empirical observations of the disorder effect in MR \cite{pudalov_0103087}.  Application of perpendicular field changes the symmetry class of the system, favors delocalization of these electrons and hence  decreases the response of a 2D  system to parallel magnetic field.
 Such explanation initiates questions about the source of 10-30\% electrons that are out of the game at $B_\perp=0$ and  about their influence on the parallel field MR.

The tilted field approach suggested in this paper might be applied for similar study of EEC with high-mobility Si-MOSFETs where strong metallic conductivity ($\partial\sigma/\partial T<0$) and metal-insulator transition are considered to be
driven by the EEC \cite{finscience}.
This task is however rather challenging experimentally because it requires a combination of millikelvin temperatures (to insure the deep diffusive limit $k_BT\tau/\hbar\ll 1$ for an order of magnitude higher mobility samples), field sweep  and sample rotation.

\section{Conclusion}
In this paper we  applied phenomenological technique of resistivity tensor analysis  to check long-standing prediction by Lee and Ramakrishnan~\cite{LeeRamakr} about Zeeman splitting dependence of the electron-electron interaction correction (EEC) to conductivity in the diffusive regime $k_BT\tau/\hbar \ll1 $. Our measurements reveal distinctly different behaviors of  the magnetotransport in the two domains of perpendicular field, the low $B_\perp$-field (LF) and the high $B_\perp$-field (HF) one.

In the LF domain, the 2D system demonstrates a strong and $T$-dependent magnetoresistance versus $B_\parallel$ field, whereas the Hall angle exceeds the Drude value by up to 30\%, depends on $B_\perp$, and grows as $T$ decreases. This Hall anomaly  obscures determination of the quantum interaction correction from the magnetotransport data  in the LF domain.

In the  HF domain, the Hall anomaly is washed-out by the $B_\perp$ field that enables  extracting the interaction quantum correction from experimental data. The EEC  was found to be field direction independent, and linear in both $\ln(T)$ and $\ln(B)$, as expected.

Remarkably, the magnitude of the experimentally determined EEC appeared to be more than a factor of 10 smaller than the parallel field magnetoconductance.  Thus,  the  parallel field MR  observed in Si-MOSFETs is not explained  by the EEC solely.
Even more surprisingly, the  observed strong parallel field MR quickly diminishes when the perpendicular field component is applied on top of the parallel field. This fact is the  direct evidence  for non-purely Zeeman origin of the parallel field magnetoresistance.

In total, our findings  point at the incompleteness of the existing theory of magnetotransport in interacting and disordered 2D systems:  too strong parallel field magnetoresistance, its suppression by perpendicular magnetic field, and the low-field Hall anomaly require an explanation. We believe that these  three phenomena are interrelated and originate from   a destructive action of the perpendicular field on the localized states.

\section{Acknowledgements}  We thank G.M.~Minkov, A.V. Germanenko, A.A. Sherstobitov,
I.S.~Burmistrov, and A.M.~Finkel'stein for discussions, and S.I.~Dorozhkin for stimulating criticism of the preliminary results. The work was supported by RFBR (grants 12-02-00579), by Russian Ministry of Education and Science (grant No 8375), and using research equipment of the Shared Facilities Center at LPI.

\section{APPENDIX A: TEST OF THE APPLICABILITY OF THE HIGH-FIELD ASYMPTOTICS}
\begin{figure}[t]
\begin{center}
\centerline{\psfig{figure=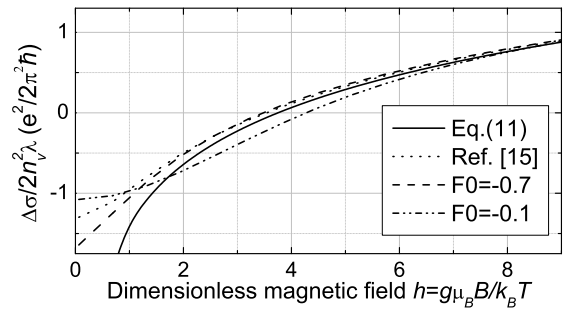, width=230pt}}
 \caption{Theoretical magnetic field dependence of the EEC, normalized by interaction
 factor.
 Solid line corresponds to high-field asymptotics Eq.~(\ref{lowTBasympt}), dotted curve corresponds to aproximation from Ref.~\protect{\cite{minkovSO}}~(see text), dashed and dash dotted curves are exact results for $F_0^{\sigma}=-0.1$ and $F_0^{\sigma}=-0.7$ respectively. The curves are shifted vertically to achieve coincidence in the  high-field limit.}
\label{asymptotics}
\end{center}
\end{figure}

In the original papers \cite{LeeRamakr,ZNABPar},  the interaction correction field dependence, $\Delta\sigma(h)$, is expressed via an integral with known low and high-field asymptotics.
This integral is inconvenient in handling,  therefore,  we used  its analytical asymptotics  Eq.~(\ref{lowTBasympt}) instead.
In Fig.~\ref{asymptotics} we compare the exact result $\Delta\sigma(h)$ for several $F_0^{\sigma}$  values  [Eq. (15) in Ref.~\cite{ZNABPar}, note that the dimensionless field scale differs by a factor $2\pi$] and high-field linear-in-$\ln(B)$ asymptotics, Eq.~(\ref{lowTBasympt}). To exclude interaction constant dependence in the high-field limit all results were divided by $2n_v^2\lambda$.
Figure~\ref{asymptotics} shows
that for $h>2$  the exact result for different $F_0^{\sigma}$ become
indistinguishable from the  logarithmic high-field asymptotics. At the same time, in the low field regime the curves substantially deviate from each other: the closer $F_0^{\sigma}$ is to -1, the stronger  is the magnetoconductance. This fact has simple physical explanation: in the high-field limit, $2n_v^2$ triplets (out of all $4n_v^2-1$ triplet terms) become suppressed, thus excluding any dependence on $F_0^{\sigma}$.
 In the low field regime, the closer $F_0^{\sigma}\equiv (2/g^*-1)$ is to -1 (i.e. to the Stoner instability), the larger is the  effective $g$-factor, and the stronger is the response of conductivity to magnetic field.

Alternative approach  used in Ref.~\cite{minkovSO} is based on
the approximation of the  crossover function (Eq.(4) from Ref.~\cite{minkovSO}).
For comparison, we present this approximation  also in Fig.~\ref{asymptotics}.

 From the practical point of view, the condition $h>2$ means that the total magnetic field $B$[T] should exceed $2.69\times T$[K]. This inequality is valid for our low-temperature high-field data shown in Figs.~\ref{15tesla} and \ref{Vg21}.

\section{APPENDIX B: LOW-FIELD NONLINEARITY OF THE HALL RESISTANCE}
Nonlinearity of the Hall resistance in the low field domain affects both $\mu(B)$ and $\Delta\sigma (B)$ dependences due to the calculation procedure, based on Eq.~\ref{aamctensor}.
Most of scattering mechanisms and quantum corrections, including low-field weak localization and Maki-Thompson corrections, renormalize mobility $\mu$ and do not renormalize Hall effect~\cite{aareview}. EEC does affect Hall resistivity but has no peculiarities close to zero field. The low field Hall feature is thus unexpected in theory, though its experimental observation is not surprising.   The lack of agreement between low-field Hall resistance and quantum correction theory was first pointed by Ovadyahu \cite{Ovadyahu} and reported several times since then \cite{HallZhang,HallNewson,Goh,MinkovHall}.  Although  there are some theoretically suggested mechanisms \cite{MinkovHall,HallFeatureTheories,Michaeli}, this phenomenon is still poorly understood.

Let us briefly summarize our experimental observations of the low-field nonlinearity in the Hall resistance in Si-MOSFETs:
(i) The effect weakens as temperature increases (see e.g. Fig.~\ref{15tesla}); (ii) The amplitude of the effect may achieve $10e^2/2\pi^2\hbar$ at low temperatures, i.e. variations in the Hall angle are as large as 30\%; (iii) The $\Delta\sigma(B)$ dependence is determined by perpendicular field component. To prove this we compare in Fig.~7  $\Delta\sigma(B)$ for perpendicular field orientation, $\Delta\sigma(B_\perp)$ and $\Delta\sigma(B)$ for $\theta=45^\circ$;
(iv) The field range, where the  Hall resistance is nonlinear, broadens as density (and hence conductivity) decreases.
(v) The effect is observed in low-mobility Si-MOSFETs ($\mu\sim 0.2$~m$^2$/Vs) in the temperature range $0.3 - 15$\,K ($k_BT\tau/\hbar =  0.01 -0.5$). For high mobility samples  ($\mu > 2$~m$^2$/Vs) we didn't observe any nonlinearity in the  Hall resistance down to 0.05K ($k_BT\tau/\hbar \approx  0.1$).

It might be that the low-field Hall nonlinearity  is somehow related to weak localization, though weak localization itself does not produce correction to Hall coefficient. Indeed, let us estimate transport field $B_{\rm tr}={\Phi_0}/{4\pi l^2}$, ($\Phi_0=(2\pi\hbar c/e)$), that is the typical value of perpendicular field component where weak localization is suppressed.
 By substituting the transport mean free path $l$ from the Drude formula
$l^{-1}= \rho_D\times(2e^2/2\pi\hbar) \times \sqrt{\pi n }$
for a two-valley system, we get: $B_{\rm tr}=n \rho_q^2\times (2\pi\hbar c/e) $, where $\rho_q\equiv \rho_{\rm D}\times (e^2/2\pi\hbar)$ is the dimensionless Drude
resistivity, and  $\rho_{\rm D}$ for simplicity was  taken equal to $\rho(B=0)$;   this simplification is justified for $\rho\ll
2\pi\hbar/e^2 \approx 26$~kOhm. For practical use $B_{\rm tr}[{\rm T}]= 0.062\times n[10^{12} {\rm cm}^{-2}]\times (\rho_D[{\rm kOhm}])^2$. The coincidence of the estimated  $B_{\rm tr}\approx 1.25$\,T value for the most intensively discussed density $n=1.25\times 10^{12}$cm$^{-2}$ with characteristic field of the Hall anomaly suppression at the lowest temperature (see Figs.~1, 3, 4) points to the relationship between weak localization and the anomalous low-field Hall slope.

 Observation (iv) is in line with this scenario, because growth of the resistivity pushes $B_{\rm tr}$ to higher fields.
A mechanism of the weak-localization-related low-field nonlinearity in the Hall resistance was suggested in Ref.~\cite{MinkovHall}, where the effect was interpreted as the second-order correction to  conductivity; i.e. a crossed term of the EEC (which controls the amplitude of the effect) and weak localization (which controls the  magnetic field dependence). In our case, however, the sign of the effect is opposite to that of Ref.~\cite{MinkovHall}, and the amplitude of the effect is much stronger.

\begin{figure}[t]
\begin{center}
\centerline{\psfig{figure=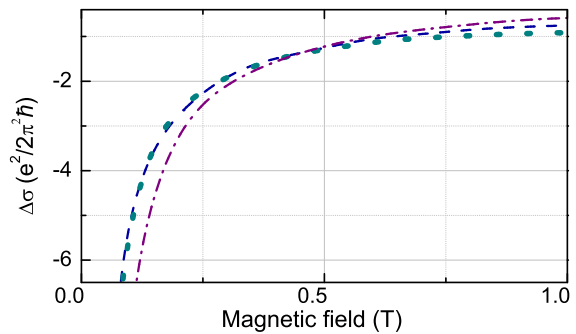, width=230pt}}
 \caption{(Color online) $\Delta\sigma(B)$ dependencies for sample Si-40, $n=1.25\times10^{12}$~cm$^{-2}$ at $T=0.6$~K for three cases: perpendicular magnetic field (dotted curve), perpendicular component of the field for the sample tilted by $45^\circ$ (dashed line, almost indistinguishable by eye from the solid curve), and total magnetic field for the sample tilted by $45^\circ$ (dash-dotted curve).
}
\label{lowfieldfeature}
\end{center}
\end{figure}

Similar sign of the effect (i.e. Hall coefficient decreasing with field down to its nominal value $1/ne$) follows from memory  effects \cite{HallFeatureTheories}. However, the memory effects do not produce temperature dependent correction to Hall effect,  as we observed.
Interaction in the Cooper (particle-particle) channel \cite{AALK} produces the so-called DoS correction that is suppressed in magnetic fields of the order of $ B\propto T/\mu$, similar to what is seen in experiment. The observed effect however is too big to be explained by quantum correction: indeed, were the Hall nonlinearity, $\Delta\sigma_c\sim10e^2/2\pi^2\hbar$,  caused by DoS correction, one would observe an enormous temperature dependence of the conductivity  (of the insulting sign) at zero field  $ \partial\sigma/\partial T \sim10e^2/2\pi^2\hbar\times \ln(T) >0$, much stronger than both, EEC and weak localization. The absence of such temperature dependence in our data points to the irrelevance of the  DoS correction  to the observed weak field Hall anomaly.

Another model that might explain field dependent Hall effect is a classical multi-component system.
However, this model can hardly be applicable to our data because  the Hall coefficient nonlinearity should be accompanied  with a large positive magnetoresistance, and should occur in fields  $\sim1/\mu$, i.e. much higher than we observe.

To conclude this discussion,   the origin of the low-field Hall anomaly currently is not clear and requires both theoretical
 reconsideration and  detailed experimental study.


\begin{thebibliography}{99}
\bibitem{commentondensity}
 The density value $n$ was determined  preliminary with $\sim 5\%$ uncertainty 
from Shubnikov-de-Haas oscillations, and adjusted more precisely to make $\Delta\sigma\approx 0$ in the high-temperature limit. Such procedure was substantiated in Ref.\protect\cite{minkovballistictodiffusion}.
\bibitem{commentonlnTEF} Original formula from Ref.\cite{ZNALarge} contains $\ln(k_BT/E_F)$ instead of $\ln(k_BT\tau/\hbar)=\ln(k_BT/E_F)+\ln(E_F\tau/\hbar)$. We omit the term $\ln(E_F\tau/\hbar)\approx 2-3$ because it is just a constant shift, which enters to mobility only and does not affect conductivity tensor in Eq.~(\ref{aamctensor}) as EEC does (for argumentation, see Ref.~\cite{minkovballistictodiffusion}).
\end{thebibliography}
\end{document}